# UPCASE - A Method for Self-Assessing the Capability of the Usability Process in Small Organizations

Thaísa C. Lacerda[a], Christiane G. von Wangenheim[a], Jean C.R. Hauck[a]

Graduate Program in Computer Science - Department of Informatics and Statistics

[a]Federal University of Santa Catarina (UFSC) - Brazil

thaisa_lacerda@hotmail.com, c.wangenheim@ufsc.br, jean.hauck@ufsc.br

## Abstract

**Context:** Designing usable products is important providing a competitive edge through user satisfaction. A first step to establish or improve a usability process is to perform a process assessment. As process assessment may be costly, an alternative for organizations seeking for lighter assessments, especially small organizations, may be self-assessments. Self-assessments can be carried out by an organization on its own to assess the capability of its process. Although there are specific assessment methods to assess the usability process, none of them provides a self-assessment method, nor has been developed considering the specific characteristics of small organizations.

**Objective:** The objective of this research is to propose a method for self-assessing the capability of the usability process in small organizations. The method consists of a usability process reference model, a measurement framework, an assessment model, and a self-assessment process supported by an online tool.

**Method:** Based on systematic mapping studies on usability capability/maturity models and software process self-assessment methods, we identified the specific requirements of such a method. The UPCASE method was systematically developed using a multi-method approach based on the ISO/IEC TR 29110 and ISO/TR 18529 standard. The method has been applied and evaluated with respect to its reliability, usability, comprehensibility and internal consistency through two series of case studies.

**Results**: The proposed method enables small organizations to assess their usability process. First results indicate that the method may be reliable. Feedback also indicates that the method is easy to use and understandable even for non-software process improvement experts.

**Conclusion**: The UPCASE method is a first step to the self-assessment of the usability process in small organizations supporting the systematic establishment and improvement of the usability process contributing to the improvement of the usability of their software products.

## 1. Introduction

Software applications nowadays are present in a diverse range of devices, such as computers, tablets, mobile phones, home appliances, etc. for numerous kinds of activities, from researching a health condition and entertainment to accessing educational resources [1]. Such changes have a significant impact on the nature of user interaction, as they offer new ways of interaction anywhere, anytime by anyone [2,3]. In this context usability becomes an important software quality attribute [2–5].

Usability is the extent to which a product can be used by specific users to achieve specific goals with effectiveness, efficiency and satisfaction in a specific context of use [6]. Usability flaws may impede the users to complete their tasks or annoy them when interaction is designed unnecessarily complex or time-consuming [7]. Furthermore, in critical contexts, such as health applications, usage errors may compromise patient safety leading to injury or even death [8]. On the other hand, investing in usability by designing software through a user-centered design process can reduce overall development cost by avoiding rework at late stages in the lifecycle [9] and speed up development [10,11]. Moreover, usability can provide a competitive edge increasing sales and retaining customers, increasing user satisfaction and software acceptance [12,13]. Thus, the question is: how to develop software applications with usability?

As any other product quality, usability is directly influenced by the software process [14,15], and, therefore, it is important to define and implement an appropriated usability process [16]. To guide the definition, implementation and improvement of software processes, typically capability/maturity models (SPCMMs), such as CMMI [17], and the ones based on ISO/IEC 15504 [18] or ISO/IEC TR 29110 [19] are used. SPCMMs aim at supporting organizations to define and continually improve their process using best practices. One way for an organization to start a software process improvement (SPI) program is to perform a process assessment in order to elicit the gap between its current practices and the ones indicated by a process reference model [20,21]. The result of such an assessment is an indicator of how well the organization's processes meet the requirements of the process reference model [22] and, thus, identifying improvement opportunities.

Besides generic SPCMMs intended to be applicable in any context, there is a trend to customize such models to target more specifically certain contexts [23]. Customized models may provide specialized support by adapting process requirements and/or providing further support for their application, for example, through low cost assessment methods or reducing the need for documentation [17,24,25]. However, considering that usability is an important software product quality characteristic, it seems that neither generic SPCMMs nor customized SPCMMs include processes specifically aiming at usability [26]. This means, that, even software organizations at the highest level of maturity seem not to be required to have established any usability process [27].

In order to solve this issue, SPCMMs focusing exclusively on usability processes were developed (such as UCDM [28], ULMM [29], UMM-P [30]) [26,31]. Most of these models are based on consolidated SPCMMs, such as the ones provided by CMMI or ISO/IEC 15504. These models propose or reference a measurement framework, but only few ones define a proper process reference model. Although these models specify high-level requirements to the usability process, they seem not to provide enough information on how to implement them in practice, which may hinder a large-scale adoption. And, although such SPCMMs are supposed to be applicable in any kind of context, it remains questionable, if they are also valid, reliable and cost efficient in current software development contexts due to a lack of application and validation of these models [26,31].

Furthermore, besides the popularity of the capability/maturity models, they are mostly applied in large organizations, not becoming popular among small organizations (with less than 50 employees) and/or agile enterprises [32]. This may be due to their detailed assessment procedure requiring considerable effort with significant costs, making their adoption often impossible for small organizations [20,33–39]. Requiring less complex and more agile assessments, lighter assessment methods are developed in form of self-assessments. Self-assessments are the most common approach to conduct a software process assessment in organizations that do not aim for certification [40]. Carrying out a self-assessment can bring many benefits when compared to other methods such as less intrusion on the organization's work routine, being a quick and inexpensive way to assess a process [38,40, 56]. As self-assessments use the organization's own human resources and are less bureaucratic, they enable a more simplified way to perform a process assessment which can be performed in a shorter period of time with fewer resources [38,42]. In addition, in organizations where capability is a new concept, self-assessments allow an easy way to improve a process [43], as they do not significantly intrude the daily routine of the organization. Self-assessment is also effective in generating an "ownership feeling" among managers regarding the process quality, as it forces them to examine their own activities [43,44].

Despite these benefits, self-assessments are not without shortcomings. Organizations using self-assessments found difficulties in planning the assessment and allocating human recurses to lead and execute them. Another difficulty is the scarcity of literature regarding the "best" approach to perform a self-assessment, as there is no guidance on which self-assessment method organizations should use [45]. Another concern when performing self-assessment is often the absence of competent assessors. As the assessors in self-assessment are not necessarily experts in software process assessment and may not be familiar with SPCMMs, there is a considerable risk of misinterpretation of the attributes to be assessed, which may impact the validity of the results of the assessment [43]. Therefore, data collection instruments used in self-assessments must be explanatory in a way that non-experts may understand the items to be measured sufficiently to correctly judge their degree of performance, e.g., preventing to wrongly considering a Gantt chart to be a project plan. Furthermore, the response scale has also to be defined carefully, as assessors in self-assessment may not have sufficient experience to correctly differentiate the degree of achievement of an item on a finer grained scale [46], e.g., deciding between partially and largely achieved. Thus, if two or more points on a scale appear to have the same meaning, respondents may be puzzled about which one to select, leaving them open to making an arbitrary choice [47]. In order to minimize the assessment effort, data collection instruments should also be comprehensive enough to measure the essential information, but at the same time be succinct enough to encourage their completion.

With a trend to develop self-assessment methods specifically for small and medium enterprises, several methods to self-assess the capability of the usability process have been developed. Most of them are based on consolidated models, such as CMMI or ISO/IEC 15504 [48]. In general, they use simplified assessment processes, focusing on data collection and analysis. Most of the methods propose to collect data through questionnaires to be answered by managers or other team members related to the process being assessed. However, so far there does not exist a self-assessment method that covers specifically the assessment of the usability process in small organizations [48].Therefore, this article present UPCASE – a self-assessment method for assessing the capability of the usability process in small organizations in accordance with ISO/IEC TR 29110.

## 2. Background

### 2.1. Usability process

Usability engineering is the application of systematic, quantifiable methods to the development of interactive software systems to achieve high quality in use [49]. It is generally concerned with human-computer interaction and specifically with the development of human-computer interfaces that have high usability or ease of use. Usability engineering provides structured methods for achieving efficiency and elegance in interface design [50], adapting the general components of software engineering to provide an engineering process to develop products with interfaces that have good usability.

This process, denominated as usability process, serves as a model for developing high usability interfaces, such as specified by ISO/TR 18529 [51] providing a formalized definition of the processes described in ISO/IEC 13407 [52], in order to make it accessible to process assessment. Figure 1 provides a general overview of the usability process reference model and its (sub-)processes [18].

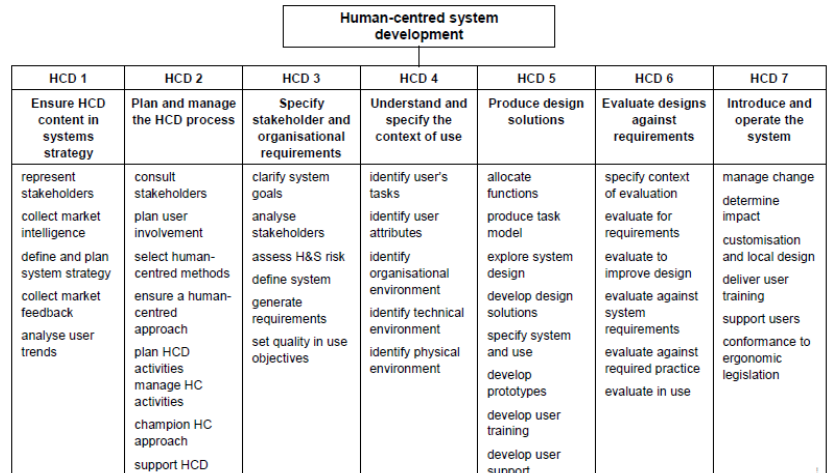

| HCD 1 | HCD 2 | HCD 3 | HCD 4 | HCD 5 | HCD 6 | HCD 7 |
|---|---|---|---|---|---|---|
| Ensure HCD content in systems strategy | Plan and manage the HCD process | Specify stakeholder and organisational requirements | Understand and specify the context of use | Produce design solutions | Evaluate designs against requirements | Introduce and operate the system |
| represent stakeholders | consult stakeholders | clarify system goals | identify user's tasks | allocate functions | specify context of evaluation | manage change |
| collect market intelligence | plan user involvement | analyse stakeholders | identify user attributes | produce task model | evaluate for requirements | determine impact |
| define and plan system strategy | select human-centred methods | assess H&S risk | identify organisational environment | explore system design | evaluate to improve design | customisation and local design |
| collect market feedback | ensure a human-centred approach | define system | identify technical environment | develop design solutions | evaluate against system requirements | training deliver user training |
| analyse user trends | plan HCD activities | generate requirements | identify physical environment | specify system and use | evaluate against required practice | support users |
| | manage HC activities | set quality in use objectives | | develop prototypes | evaluate in use | conformance to ergonomic legislation |
| | champion HC approach | | | develop user training | | |
| | support HCD | | | develop user support | | |

**Figure 1. Human centered design process (Source: (ISO/IEC, 2000))**

## 2.2. Process Improvement and Assessment

In any process improvement program, it is essential to identify the organization's real problems and improvement opportunities that effectively bring benefits. In this way, process assessment is among the first activities when starting a process improvement program [53]. Process assessment is a disciplined assessment of the processes of an organization against a process reference model compatible with a process assessment model [18]. ISO/IEC 15504 [18] presents an assessment framework that defines the elements necessary to carry out a process assessment. Following ISO/IEC 15504, the framework for conducting assessments, called assessment method, includes a process reference model, a measurement framework, an assessment model, as well as an assessment process. CMMI, on the other hand, defines the references model as part of the assessment model [17].

The process reference model describes a process life cycle, defining its purposes, process outputs and the relationships between them [18]. Typically, reference models are refined into activities [54] or base practices [55] that should be carried out, so the process might achieve its goal. Processes can be grouped into process areas, a group of related practices that when implemented satisfy important goals for making improvement in that area. Sometimes process areas are presented as dimensions or categories, which represent key elements of the process [17].

The measurement framework provides a base for rating the capability of processes and/or the maturity of the organization, based on their achievement of defined process attributes. It typically includes: process attributes, a rating scale and a capability/maturity scale [18]. A process attribute represents measurable characteristics, which support the achievement of the process purpose and contribute to meeting the business goals of the organization [18]. A rating scale typically is an ordinal scale to measure the extent of the achievement of a process. The capability scale is composed by capability levels and represents the capability of the implemented process in increasing order, from not achieving the process purpose, to meeting current and projected business goals [17]. The maturity scale characterizes the maturity of the organization and each level builds on the maturity of the level below (Figure 2). Capability and maturity levels are typically represented using a staged or continuously scale system. The continuous representation uses capability levels to characterize capability relative to an individual process area. A staged scale represents the maturity level of the organization's processes. Each maturity level is comprised of a set of process areas. To reach a certain maturity level, the set of process areas must met a certain capability level [17].

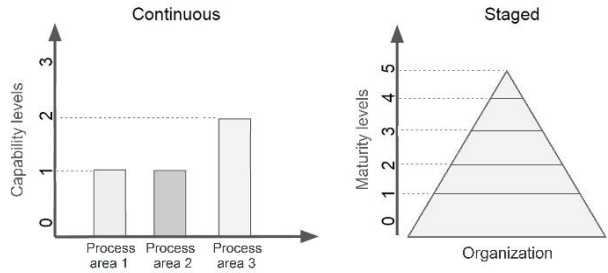

Figure 2. Examples of continuous and staged scale (adapted from SEI (2010))

The assessment process is a set of activities that must be performed to conduct an assessment [56]. It contains activities such as planning, data collection, data validation, process attribute rating and reporting [18], defining also their inputs and outputs (Figure 3). Each activity can be performed by adopting specific techniques (such as interviews, workshops, meeting with stakeholders, presentations) and using specific tools (such as spreadsheets, templates or software systems).

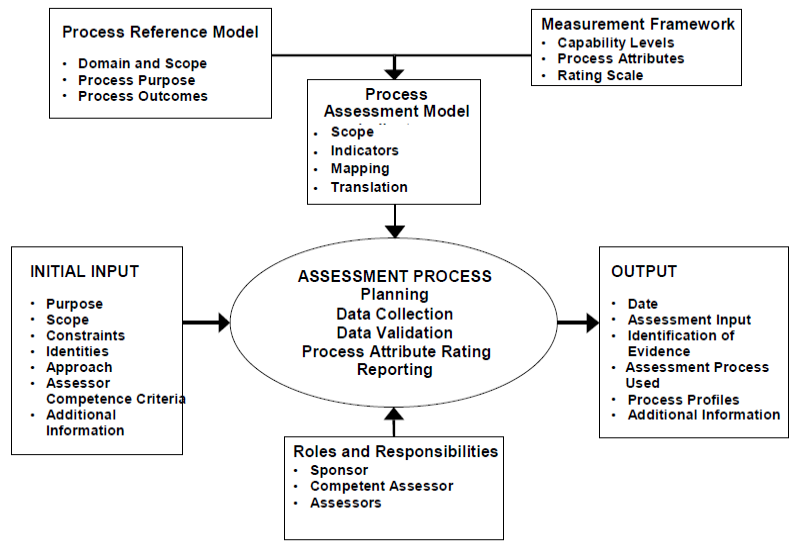

Figure 3. Components of the assessment framework (ISO/IEC 2004)

There are basically two ways for performing process assessments: as an independent assessment performed by a team external to the organization, or as a self-assessment performed by a team internal to the organization being assessed [18]. Self-assessment is performed by an organization to assess the capability of its own process. Assessors of a self-assessment are usually member of the organization [57]. A convenient way to conduct an assessment is to interview staff who executes the process and review related documentation. The assessment result is usually reserved for internal use in order to support process improvement [41].

## 3. Methodology

We adopted a multi-method approach for this research as presented in Figure 4.

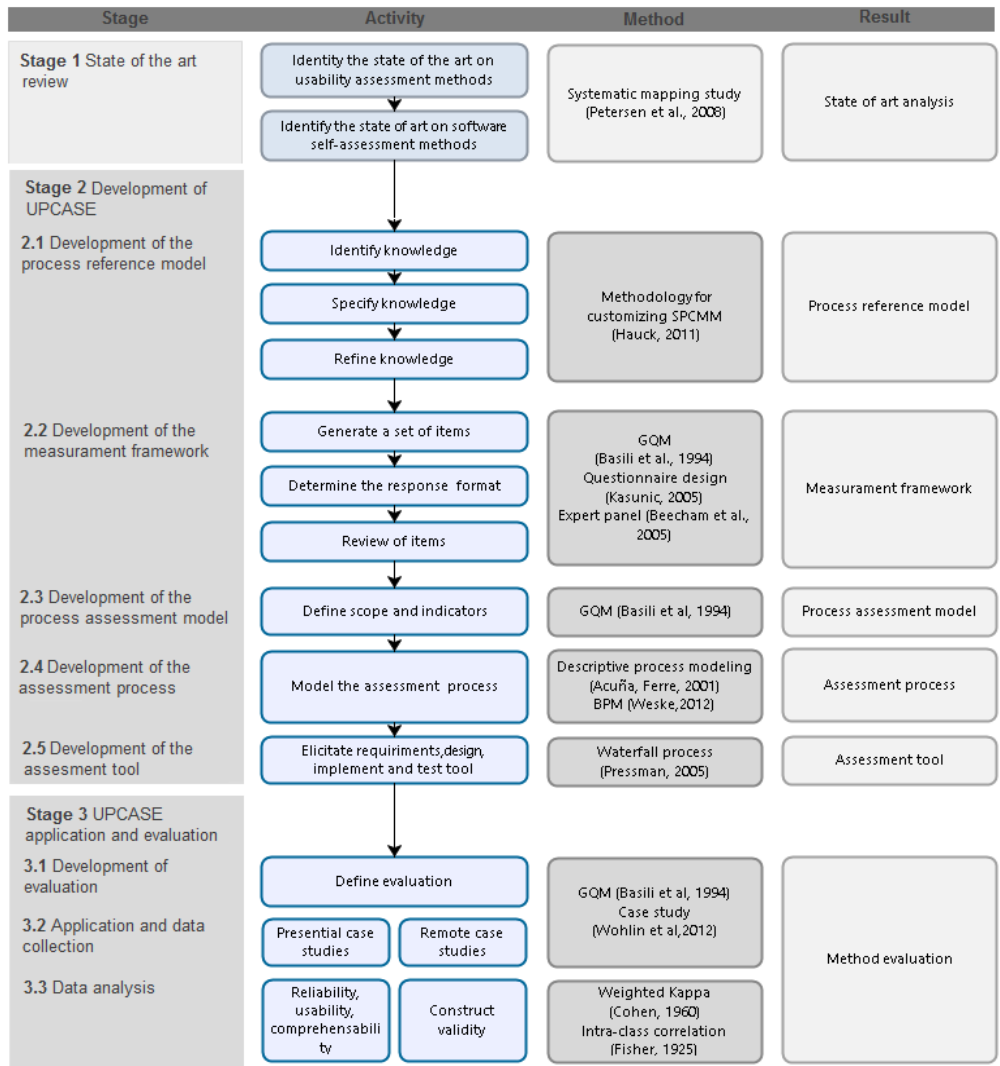

**Figure 4. Research methodology**

Initially, we preformed systematic mapping studies following the procedure proposed by Petersen [58] in order to synthesize the state of the art with respect to existing usability maturity/capability models [26] as well as software processes self-assessment methods [48]. Based on these overviews, we started the development of the usability capability self-assessment method by analyzing the context and eliciting and analyzing requirements with respect to the method. Based on the standard ISO/IEC TR 29110 [19], we modeled the structure and elements of the method. The development of the usability process self-assessment method has then been done in several steps:

**2.1 Development of the process reference model.** The reference model is developed based on the state of the art following the methodology for customizing SPCMM as proposed by Hauck [59]:

- Knowledge identification aiming at the familiarization with the domain and the characterization of the context for which the assessment method is being customized, defining its scope and objectives.
- Knowledge specification developing a first version of process reference model.
- Knowledge refinement evaluating and consolidating the process reference model.

**2.2 Development of the measurement framework.** The measurement framework defines an ordinal scale for the assessment of the process capability. It is developed by adopting the Goal/Question/Metric (GQM) approach [60] and following the questionnaire design guide proposed by Kasunic [61]:

- Generation of a set of items by systematically decomposing the process reference model into questionnaire items.
- Determination of the response format for the data collection instrument items based on existing measurement frameworks and taking into consideration requirements for self-assessment methods.
- Evaluation of the face validity of the questionnaire items through an expert panel as proposed by Beecham [62] using the feedback from the experts to improve the draft model.

**2.3 Development of the process assessment model.** The process assessment model is developed in compliance with ISO/IEC TR 29110-3 and defines the scope and indicators based on the UPCASE process reference model and measurement framework.

**2.4 Development of the self-assessment process** based on the process of empirical studies as proposed by Wohlin [63]. The process is modeled in a prescriptive way defining how the self-assessment process should be executed [64]. The self-assessment process is represented by using the Business Process Modeling Notation (BPMN) [65].

**2.5 Development of an assessment tool** in order to support the usage of the self-assessment process. The tool is developed following a waterfall development process [66]:
- Requirements analysis: based on the defined process, functional and non-functional requirements of the tool are defined and use cases are described.
- Design: the architecture of the tool is modeled, and the design of the user interfaces is prototyped.
- Implementation: the tool is constructed as a web-application.
- Test: throughout its development the tool is being tested on different levels, including system tests to ensure that all requirements have been met.

We, then, applied and evaluated the usability process self-assessment method in terms of reliability, usability, comprehensibility and internal consistency:

**3.1 Definition of the evaluation.** The evaluation is systematically defined using GQM through a series of case studies [63], in which small organizations applied the UPCASE method in order to assess the capability of their usability process. As part of this step the evaluation objective, measures and data collection has been defined.

**3.2 Application and data collection.** We performed two types of cases studies: unobserved case studies, in which the assessment has been performed remotely without the observation of a researcher collecting data via the UPCASE tool; and observed case studies, in which the assessment has been performed by the organizations themselves, with a researcher observing, who also performed an independent assessment.

**3.3 Data analysis.** To analyze reliability, we performed a concordance analysis using the Weighted Kappa coefficient [67], as well as the analysis of the intraclass correlation coefficient [68]. The intraclass correlation coefficient is often applied for assessing the consistency or reproducibility of quantitative measurements made by different observers measuring the same quantity [68]. However, considering only the results of the evaluations, this coefficient does not allow to evaluate agreement. Weighted Kappa coefficient, on the other hand, is a widely used index for assessing agreement between raters. When the rated categories are ordered or ranked, then a weighted kappa coefficient is computed, considering the different levels of disagreement between categories. Usability and comprehensibility of the UPCASE method were analyzed based on the qualitative feedback from the participants of the case studies and the observations. The results of the self-assessments and the researcher's assessment are compared to analyze the internal consistency in UPCASE questionnaire. Internal consistency was analyzed using Cronbach's alpha [69].

## 4. Requirements to a Self-Assessment Method for Assessing the Usability Process in Small Organizations

In addition to the small number of employees, small organizations have other characteristics that make their needs and way of working unique. Small organizations often do not have enough staff to develop specialized functions that would enable them to perform complex tasks or develop secondary products [70,71]. Also, in general these organizations do not use formal processes and, therefore, may have many problems to complete their projects under time and cost constraints [72]. The limitation of financial resources presents many consequences, as it hinders the hiring of specialists and the training of employees; and complicates the execution of processes improvement programs, which typically require a considerable amount of time and money [71–74]. These particular characteristics of small organizations directly impact on how they assess and improve their software processes. As a consequence their main motivation to implement software process improvement may not be to obtain a certification, but rather to make the organization's process more efficient and effective [70,75].

These characteristics may lead small organizations to use light assessment methods that can be used with the resources (human and non-human) that are available, carrying out the assessment within reasonable time with low cost [42,70,71,73]. Methods that are publicly available and being supported by tools that (semi-)automate and/or support assessment steps are preferred [73,74,76]. The lack of specialized SPI professionals in small organizations makes it necessary to use methods that can be used by non-experts [70,76]. Therefore, it is important that the assessment method provides access to a detailed definition of the process reference model and the assessment model, with descriptions of process purpose, process outcomes, capability levels and process attributes [19]. This also applies in relation to the usability process, as small organizations generally present a low level of maturity and little knowledge about relevant usability concepts and standards [77–80]. Another reason that makes process assessment less attractive to small organizations is the difficulty of understanding and implementing them in practice [36,81]. This fact leads many organizations to seek even more for simplicity of processes and as result they are increasingly attracted to agile methods. Considering this, those agile approaches should be incorporated into any potential process assessment method [20].

In order to develop an effective and efficient method for self-assessing the usability processes in small organizations, we elicitated a set of requirements (Table 1) based on requirements for software/usability process assessment methods, self-assessment and characteristics of small organizations found in literature.

| No. | Requirement | Element | Source(s) |
|---|---|---|---|
| 1 | The method should allow a fast-internal assessment. | Method | [42,72,73,79] |
| 2 | The method should allow getting a snapshot of actual processes. | Method | [73] |
| 3 | The method should be of low cost. | Method | [20,42,72–74,76] |
| 4 | The method should provide the necessary tools (including tools for (partial) automation, eliminating laborious manual work and extensive documentation). | Method | [20,42,73,74,76] |
| 5 | The method should be based on already established SPI standards that are widely recognized. | Method | [20,42,70,76,82] |
| 6 | The method should not require staff to have prior SPI experience, specific software engineering knowledge nor require the involvement of external experts. | Method | [19,20,42,70,76] |
| 7 | The method should provide accesses to a detailed definition of the process reference model and the assessment model, with descriptions of process purpose, process outcomes provided by the PRM and capability levels and process attributes. The rating scale needs to be supported by a comprehensive set of indicators of process performance. | Method | [20,42,76] |
| 8 | The method should be public available. | Method | [20,42,76] |
| 9 | The method should support the identification of improvement suggestions. | Method | [20,42,76] |
| 10 | The process assessment should guide the activities that need to be performed in an assessment. It should provide a clear definition of roles and their responsibilities and a detailed description of the self-assessment process, with recommendations that are easy to understand. | Assessment Process | [19,20,42,73,74,76,80] |
| 11 | The process assessment should require few resources. | Assessment Process | [72,74] |
| 12 | The process assessment should consider the views of the team while indicating what needs to be improved. | Assessment Process | [70] |
| 13 | The process assessment and measurement framework should facilitate | Assessment Process / | [39,70] |

| | self-assessment. | Measurement framework | |
|---|---|---|---|
| 14 | The measurement framework should provide a basis for use in process improvement and capability determination. | Measurement framework | [39] |
| 15 | The measurement framework should take into account the context in which the assessed process is implemented. | Measurement framework | [39] |
| 16 | The measurement framework should contain a process capability scale. | Measurement framework | [39] |
| 17 | The measurement framework should be applicable across all application domains mainly for very small entities. | Measurement framework | [39] |
| 18 | The measurement framework should provide an objective benchmark between organizations. | Measurement framework | [39] |
| 19 | Processes should be light, easily implementable, representing well-focused life cycle profiles, not requiring processes that do not make sense. | Process Reference Model | [73,74,83] |
| 20 | Processes should avoid complex nomenclature, concepts and practices. | Process Reference Model | [19,42,77–80] |
| 21 | Processes should be strongly human oriented and emphasizing communication performed face to face. | Process Reference Model | [19,77] |
| 22 | Processes should focus on the Engineering Process group. | Process Reference Model | [19,42] |
| 23 | Processes should aim at involving user in the usability lifecycle. | Process Reference Model | [84,85] |
| 24 | Processes should not impose rigorous and inflexible methods and practices. | Process Reference Model | [84] |
| 25 | Practices should be simple. | Process Reference Model | [85] |
| 26 | Processes should be flexible and allow iteration. | Process Reference Model | [79] |

**Table 1. Requirements to a self-assessment method for assessing the usability process in small enterprises**

## 5. UPCASE – A Method for Self-assessing the Capability of the Usability Process in Small Organizations

Based on the state of the art and with respect to the identified requirements, we developed UPCASE, a method for self-assessing the capability of the usability process in small organizations in alignment with ISO/IEC TR 29110-3 [39]. The scope of the UPCASE method covers the first objective of ISO/IEC TR 29110-3-1, focusing on "Assessing the process capability based on a two-dimensional evaluation model, containing a process dimension and a quality dimension of the process". The original ISO/IEC TR 29110-1 elements have been simplified, so that UPCASE uses only the elements necessary to achieve the first objective of the standard (Figure 5).

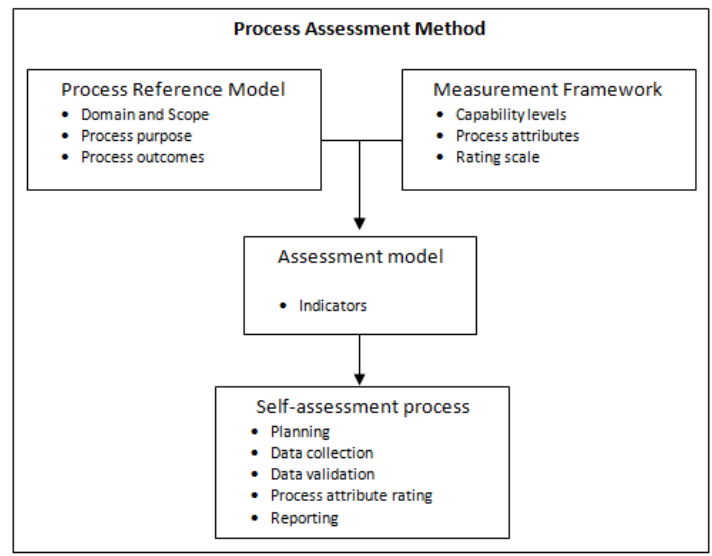

**Figure 5. Elements of the UPCASE assessment method (adapted from ISO/IEC TR 29110-3-1 (2015))**

## 5.1 UPCASE Measurement Framework

The measurement framework provides a schema to be used to characterize the capability of a process in relation to a reference model. The measurement framework of the UPCASE Method is based on ISO/IEC TR 29110 and is composed of three elements: capability levels, process attributes and a rating scale. Capability levels are used to determine the process capability. Capability levels group the process attributes and define an ordinal scale of capability that is applicable across all processes. Considering the predominance of lower capability levels in small organizations, only the first two capability levels are considered (Table 2).

| Capability level | Description |
|---|---|
| Level 0: Incomplete process | The process is not implemented or fails to achieve its process purpose. At this level there is little or no evidence of any systematic achievement of the process purpose. |
| Level 1: Performed process | The implemented process achieves its process purpose. |

Table 2. Process capability level description (ISO/IEC TR 29110)

Process attributes (PA) are measurable characteristics of the process capability that are applicable to any process. No process attributes are allocated to Capability level 0, characterizing an incomplete implementation of the process. Capability level 1 is characterized by one process attribute PA 1 Performance. This attribute is a measure of the extent to which the process purpose is achieved. The rating scale is a defined ordinal scale of measurement used to measure the extent of achievement of a process attribute, adapted from ISO/IEC TR 29110 by unifying the levels Partially achieved and level Largely achieved in order to simplify the rating process, making it easier for assessors to differentiate between rating levels [46]. Thus, UPCASE's rating scale is composed of the levels: N- Not achieved 0 to 15% achievement; P- Partially achieved >16% to 85% achievement; F- Fully achieved >86% to 100% achievement. The process profile (to determine the process capability level) is generated as defined by ISO/IEC 15504 and ISO/IEC TR 29110. The percentage of the process attribute achievement (PPAA) is calculated based on the process indicators rating:

Formula 1: PPAA = ($\sum$ process indicator rating / n° process indicators * 2)*100.

## 5.2 UPCASE Process Reference Model

The purpose of the Process Reference Model (PRM) is to define the usability process in terms of purpose and outcomes. Again, we defined the PRM in compliance with the structural definition given by ISO/IEC TR 29110-3 specifying the process purpose and process outcomes. Considering that small organizations typically need simpler processes (REQ 19) in the context of small and less complex projects [86], we based the development of UPCASE's PRM on ISO/TR 18529, since it focuses on technical process (REQ 22). As result, the UPCASE PRM includes four categories as defined in ISO/TR 18529. Taking into consideration the requirements identified in Section 4, further adaptations have been done to customize the PRM to the specific characteristics of small organizations. Considering REQ 22, 3 processes from ISO/TR 18529 were not considered: HCD 1, HCD2 and HCD 7. HCD 1 and HCD 2 were excluded, as they deal mainly with management and business strategy practices, not focusing on technical practices. On the other hand, the HCD 7 process was excluded, as it deals with the implementation and support of the system practices that are generally not a responsibility of a small organization.

In total, 10 outcomes from ISO/TR 18529 were excluded. These outcomes have been removed as they are typically not in the scope of the processes of small software organizations (such as the responsibility of installing and operating the system) or require more advanced usability knowledge than the staff of small organizations usually has. The justification for excluding the outcomes is presented in the column Justification (marked in red) in Table 3, presenting separately the outcomes with respect to each of the processes.

| Purpose | Outcome | Justification for exclusion |
|---|---|---|
| UP1 Specify stakeholder and organizational requirements | | |
| To establish the requirements of the organization and other interested parties for the system. This process takes full account of the needs, competencies and working environment of each relevant stakeholder in the system. | Required performance of the new system regarding its operational and functional objectives. | |
| | Relevant statutory or legislative usability requirements, depending on the system domain. | |
| | Co-operation and communication between users and other relevant parties | |
| | The users' jobs (including the allocation of tasks, users' comfort, safety, health and motivation) | This outcome overlaps with the outcomes "Definition of the characteristics of the intended users" and "Definition and characterization of the tasks the users are to perform" from UP2. |
| | Task performance of the user when supported by the system | |
| | Work design, and social practices and structure | This outcome overlaps with the outcome "Definition and characterization of the tasks the users are to perform" from UP2. |
| | Feasibility of operation and maintenance | REQ 22 (Process should focus on engineering process) |
| | Objectives for the operation and/or use of the software and hardware components of the system. | |
| **UP2 Understand and specify the context of use** | | |
| To identify, clarify and record the characteristics of the stakeholders, their tasks and the social and physical environment in which the system will operate. | Definition of the characteristics of the intended users. | |
| | Definition and characterization of the tasks the users are to perform. | |
| | Definition and characterization of the social and environment in which the system is used. | |
| | Definition and characterization of the technical environment in which the system is used. | |
| | The use of context analysis results in requirements to the interface design. | |
| | The context of use is available and used at all relevant points in the system development. | |
| | Definition of the characteristics of the intended users. | |
| **UP3 Produce design solutions** | | |
| To create potential design solutions by drawing on established state-of-the-art practice, the experience and knowledge of the users and the results of the context of use analysis. | Results of socio-technical context of use analysis are considered in the design. | |
| | User characteristics and needs will be taken into account in the purchasing of system components. | There are no mandatory or optional requirements related to the Acquisition processes in the Basic Profile of the ISO/IEC TR 29110 series; Nor are they expected to be defined in the next ISO/IEC TR 29110 series profiles [87]. |
| | Results of the user analysis are taken into account in the design of the system. | |
| | Existing knowledge of best practice from socio-technical systems engineering, ergonomics, psychology. | REQ 19 (Little awareness on usability) Small organizations typically does not have HR with expertise in usability |
| | Cognitive science and other relevant disciplines will be integrated into the system. | REQ 19 (Little awareness on usability) Small organizations typically does not have HR with expertise in usability |
| | Communication between stakeholders is improved because the design decisions are more explicit. | |
| | The development team is able to explore several design concepts before they settle on one. | |
| | Feedback from end users and other stakeholders is incorporated in the design early in the development process. | |
| | It is possible to evaluate several iterations of a design and alternative designs. | |
| | The user's tasks are analyzed in relation to their, navigation, hierarchy and information architecture. | |
| | The design of all the user-related components of the system is specified, in terms of "look and feel". | |
| | The interface between the user and the software, hardware and organizational components of the system are designed. | |
| | User training and support will be developed. | ISO/IEC TR 29110-4 (Small enterprises generally are not responsible for the management, operation, integration and installation of the system.) |
| **UP4 Evaluate designs against requirements** | | |
| To collect feedback on the developing design. This feedback will be collected from end users and other representative sources. | Feedback is provided to improve the design. | |
| | There is an assessment of whether stakeholder and organizational usability objectives have been achieved or not. | |
| | Long-term use of the system will be monitored | ISO/IEC TR 29110-4 (Small enterprises generally are not responsible for the management, operation, integration and installation of the system.) |
| | Potential problems and scope for improvements in: the technology, supporting material and social or physical environment. | |
| | Which design option best fits the functional and stakeholder and organizational requirements. | |
| | Feedback and further requirements from the users. | This outcome overlaps with the outcome "Feedback is provided to improve the design" from UP4. |
| | How well the system meets its organizational goals. | This outcome overlaps with the outcome "There is an assessment of whether stakeholder and organizational usability objectives have been achieved or not" from UP4. |
| | Guarantee that a particular design meets the human-centered requirements. | |
| | Conformity to international, national and/or statutory requirements, depending on the system domain. | |

Table 3. Usability process' purposes and outcomes

## 5.3 UPCASE Process Assessment Model

The UPCASE Process Assessment Model (PAM) is compliant with ISO/IEC TR 29110-3 [19] and contains the basis for collecting evidence and rating process capability. It contains the Process Dimension, which defines the set of processes that will be assessed (defined in the PRM) and the Capability Dimension, which defines the capabilities related to each process capability level and each process attribute. It defines the scope, indicators and a mapping for a Process Reference Model and a Measurement Framework.

Again, in accordance to the identified requirements and the process defined by ISO/IEC TR 29110-4, some practices of ISO/TR 18529 have been excluded or adapted to meet the requirements of the self-assessment method in this specific context. The adaptation of the practices aims at meeting requirements 6, 19, 20 and 25 identified in Section 4. Therefore, the practices are written in such a way that staff without SPI or usability knowledge can understand them. To accomplish this, the use of complex nomenclature and concepts and jargons from the usability domain was avoided. Furthermore, for each of the work products an example is provided, illustrating the expected result. Aiming at a "light" process, practices that overlap each other or that seem to complex in the context of small organizations were removed. The customization of the practices is presented in Table 4.

| Id. | ISO/TR 18529 practices | Customized UPCASE practices | Justification for exclusion |
|---|---|---|---|
| **UP1 Specify stakeholder and organizational requirements** | | | |
| 1 | Clarify system goals | Identify system purpose. | |
| - | Analyze stakeholders | -- | This practice overlaps with "Identify and document significant user attributes" practice. In addition, the basic profile of ISO/IEC TR 29110-4 does not have any practice related to the analysis of the roles of each stakeholder group |
| | | | besides the users. Characterization of the users is covered through UP2-Practice 6. |
| - | Assess H&S risk | -- | This practice has been removed in order to keep the process simple (REQ 19), and because it is contained in practice 6. |
| 2 | Define system | Define system performance and behavior requirements desired by the user. | |
| - | Generate requirements | -- | This practice is performed in the context of the software engineering process (ISO/IEC 12207). Its output, however, should be used as input in the usability process, being necessary for the execution of practices 3 and 4. |
| 3 | Set quality in use objectives | Define usability requirements. | |
| **UP2 Understand and specify the context of use** | | | |
| 4 | Identify and document user's tasks | Identify and describe the user's tasks of the system. | |
| 5 | Identify and document significant user attributes | Identify user characteristics. | |
| 6 | Identify and document organizational environment | Identify social environment characteristics. | |
| 7 | Identify and document technical environment | Identify device characteristics. | |
| 8 | Identify and document physical environment | Identify physical environment characteristics. | |
| - | Allocate functions | -- | This practice has been removed in order to keep the process simple (REQ 19), and because it is contained in practice 10. |
| **UP3 Produce design solutions** | | | |
| 9 | Produce composite task model | Analyze user's tasks. | |
| 10 | Explore system design | Develop and analyze design options during interface development. | |
| 11 | Use existing knowledge to develop design solutions | Develop design solutions using existing knowledge. | |
| 12 | Specify system and use | Specify all user-related elements of the system. | |
| 13 | Develop prototypes | Prototype all user-related elements of the system. | |
| - | Develop user training | -- | ISO/IEC TR 29110-4 (Small enterprises generally are not responsible for the management, operation, integration and installation of the system.) |
| - | Develop user support | -- | ISO/IEC TR 29110-4 (Small enterprises generally are not responsible for the management, operation, integration and installation of the system.) |
| **UP4 Evaluate designs against requirements** | | | |
| 14 | Specify and validate context of evaluation | Prepare prototype/system evaluation. | |
| - | Evaluate early prototypes in order to define the requirements for the system | -- | This practice has been removed in order to keep the process simple (REQ 19), and because it might be contained in practice 15. |
| 15 | Evaluate prototypes and in order to improve the design | Evaluate prototypes and system to find usability problems. | |
| 16 | Evaluate the system in order to check that the stakeholder and organizational requirements have been met | Evaluate system against usability requirements | |
| - | Evaluate the system in order to check that the required practice has been followed | Evaluate system to find usability problems. | This practice has been removed in order to keep the process simple (REQ 19), and because it might be contained in practice 15. |
| - | Evaluate the system in use in order to ensure that it | -- | ISO/IEC TR 29110-4 (Small enterprises generally are not responsible for the |

| | | | management, operation, integration and installation of the system) |
|---|---|---|---|
| continues to meet organizational and user needs | | | |

**Table 4. UPCASE process practices**

In order to facilitate the understanding of its practices, ISO/TR 18529 provides a description for each of them (Table 5). These descriptions have been adapted in order to attend REQ 20, helping assessors to better understand the UPCASE practices.

In order to better assist in the correct implementation of the usability process assessment, as well as to help the assessment team to verify if the usability practices of the SE are in accordance with the UPCASE assessment method, the UPCASE method provides additional artifacts, such as the description of the self-assessment process and the assessment questionnaire. The assessment questionnaire should be used during the assessment as a "roadmap", which allows the assessment team to judge each practice of the UPCASE PRM. Therefore, the questionnaire includes one item to assess each of the practices of the usability process. To support the judgment of the performance of each practice of the usability process, the questionnaire presents an indicator for each of them. The indicators objectively demonstrate characteristics of the practices of the assessed process. Table 5 presents examples of the indicators, a complete definition is given in [48].

| Id. | Practice | Description | Indicator | Example of techniques | Example of work products |
|---|---|---|---|---|---|
| **UP1 Specify stakeholder and organizational requirements** | | | | | |
| 1 | Identify system purpose | Identify and describe the purpose of the system, this is, the objective(s) that the user wants to achieve using the system. | Our team identifies and describes the purpose of the system. | Survey, interview, observation. | Purpose(s) of the system |
| 2 | Define system performance and behavior requirements desired by the user. | Identify the stakeholder's requirements regarding the behavior and performance of the system. The requirements cover each aspect of the system related to its use and its interface in a context of use. | Our team identifies system performance and behavior requirements desired by the user. | Survey, brainstorming, interview, observation. | System performance and behavior requirements desired by the user. |
| 3 | Define usability requirements. | Define an explicitly statement for each usability requirements, regarding its effectiveness, efficiency and user satisfaction based on the context of use analysis. The statements should be measurable objectives. | Our team defines explicit statements of usability requirements based on the context analysis. | Benchmarking with concurrent systems, synchronic analyzes formal work analyses. | A list of usability requirements |
| … | … | … | … | … | … |

**Table 5. Examples of practices descriptions and indicators**

Furthermore, suggestions of techniques and work-products are given for each indicator based on ISO/TR 18529, as well as a glossary.

## 5.4 UPCASE Self-Assessment Process

The purpose of the assessment process is to systematically guide the process assessment activities. The self-assessment process of the UPCASE method is based on the assessment process defined by ISO/IEC TR 29110-3, composed in four phases: Plan the assessment, Collect and validate the data, Generate results and Report the assessment, as presented in Figure 6. Activities that may be automated by UPCASE Tool are presented in yellow. The definition of the techniques and tools to perform each of these phases is based on good practices identified in literature.

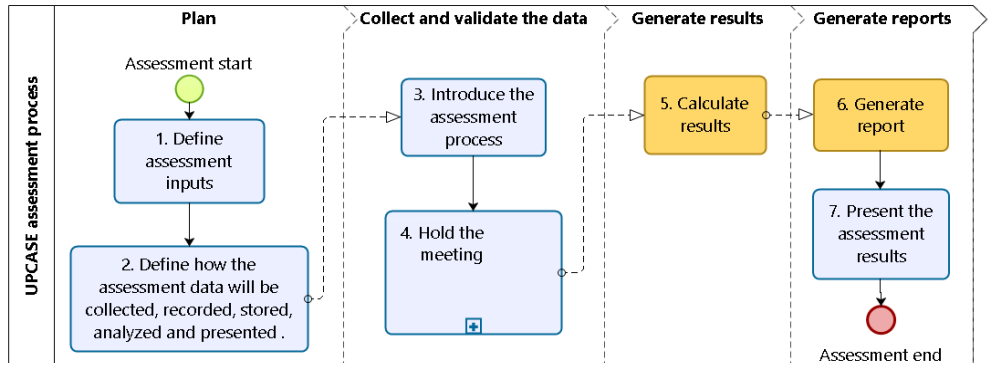

**Figure 6. Self-Assessment process**

**Plan the assessment.** During this meeting the assessment plan as pre-defined by UPCASE should be revised and the resources, schedule and roles & responsibilities should be defined. The activities of this phase can be carried out during a meeting with some members of the organization, who are responsible for the usability process, as defined in the input Roles and Responsibilities.

| UPCASE Role | Responsibilities | Role description |
|---|---|---|
| Sponsor | a) verify that the individual who is to take responsibility for conformity of the assessment is a competent assessor (following the definition of "moderator" as given by the Inputs of UPCASE);<br>b) ensure that resources are made available to conduct the assessment;<br>c) ensure that the assessment team has access to the relevant resources. | Some leadership position of the organization that realizes the need to assess the usability process, such as: Project manager, Development leader, UI Design leader. |
| Moderator | a) confirm the sponsor's commitment to proceed with the assessment;<br>b) ensure that the assessment is conducted in accordance with the assessment method;<br>c) ensure that participants in the assessment are briefed on the purpose, scope and approach of the assessment;<br>d) ensure that all members of the assessment team have knowledge and skills appropriate to their roles;<br>e) ensure that all members of the assessment team have access to appropriate documented guidance on how to perform the defined assessment activities;<br>f) ensure that all assessors are able to participate in the assessment meeting.<br>g) carry out assigned activities associated with the assessment, e.g. detailed planning, data collection &validation and reporting; | Should be chosen by the sponsor. Preferably should be a professional with more knowledge about process assessment or usability. |
| Assessor | a) provide examples of work products and techniques as evidence of the execution of the process.<br>b) rate the processes attributes. | Assessors may be any employee who performs activities related to the usability process, such as designers, system analysts, testers, etc. |

**Table 6. Definition of roles and responsibilities**

**Collect and validate the data.** During this phase the moderator presents the purpose of the process assessment. S/he presents the focus group methodology (assessment poker) and the self-assessment process, as well as the inputs and expected outputs of the assessment. This activity is supported through a script. Then, data collection and validation are performed during an assessment meeting as illustrated in Figure 7. To conduct the assessment meeting, UPCASE provides a questionnaire that contains the items that should be assessed for each usability process, as well as the description of each of them with examples of work products and techniques. UPCASE also provides a deck of assessment poker cards representing the rating scale through 3 cards ("Not achieved", "Partially achieved", "Fully achieved").

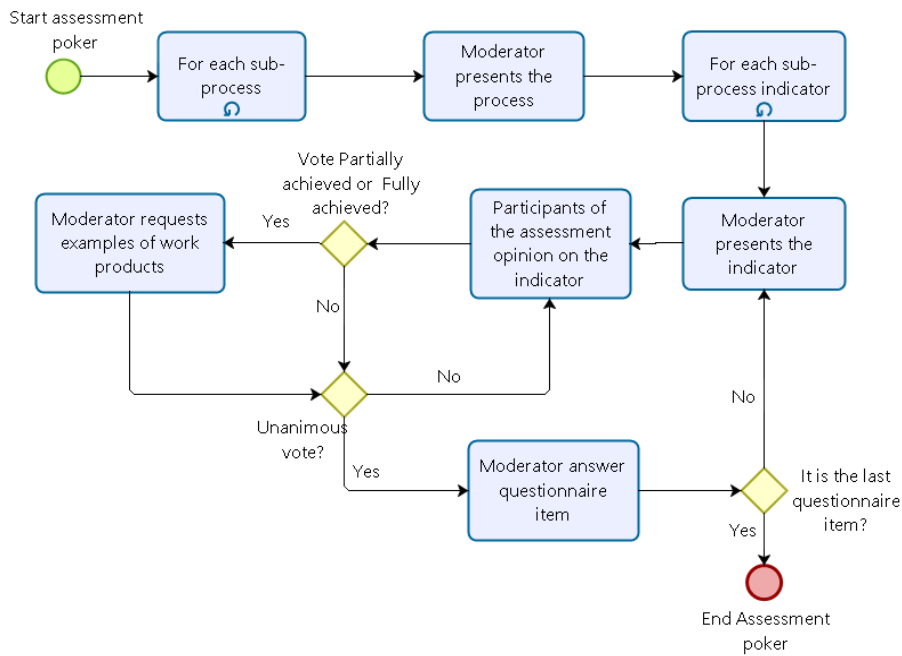

**Figure 7. Assessment meeting activities**

During Assessment Poker, similarly to Planning Poker [88], the moderator presents an indicator from the UPCASE questionnaire, and, if necessary, consults further information (such as the glossary, examples of techniques and/or work products) for clarification. Then, each participant of the assessment team rates the degree of achievement of the indicator by choosing the respective card. All participants at once turn over their card. In case of deviation of opinion, the assessors briefly justify their different opinions. Then again, each assessor expresses his/her opinion by selecting a card and turning them over, repeating this process until consensus is achieved. In order to validate the rating, the moderator may request examples of work products that demonstrate the achievement of the practice indicator from the participants. Once consensus is achieved, the moderator completes the response of the item in the UPCASE questionnaire.

**Generate results.** After the assessment meeting the assessment results are generated based on the answers of the completed UPCASE questionnaire. The capability level and the profile of the usability process are derived from the process attribute ratings by calculating the usability sub-process percentage of achievement (USPA) based on the indicator ratings:

Formula 2: USPA= $((\sum \text{sub} - \text{process indicator ratings}) /\text{n}°\text{indicators}*2)*100$.

Then the sub-process attribute capability rating is calculated based on its achievement percentage, as defined in ISO/IEC TR 29110 (Section 5.1). The usability process percentage of achievement (UPPA) is calculated based on the indicators ratings:

Formula 2:UPPA= $((\sum \text{usability process indicators ratings}) /\text{n}°\text{indicators}*2)*100$.

**Report assessment results.** Following the template provided by UPCASE, the report should summarize the assessment findings, the process profile, observed strengths and weaknesses and potential improvement actions. The report can also be generated automatically using the online UPCASE tool. Otherwise, the report may be prepared by any member of the organization that participated in the assessment. The assessment results are then presented during a meeting. The emphasis of the presentation should not be on the process rating, but rather on the items identified as improvement opportunities. During the meeting improvements actions may be discussed based on the assessment results.

The detailed description of the UPCASE self-assessment process and artefacts is given in [48].

## 5.5 UPCASE Tool

In order to facilitate the self-assessment using the UPCASE method, a web–based application, called UPCASE Tool, was developed. The tool provides support for the performance of the assessment supporting the collection of responses to the UPCASE questionnaire, gives access to additional information (such as, glossary, work products, etc.) as well as automatically generates the assessment results and report.

**Figure 8. Example screenshots of the UPCASE tool**

The tool has been implemented to work with the PostgreSQL Database Manager System (DBMS), and was programmed in PHP and JavaScript languages. The UPCASE Tool is available online http://match.inf.ufsc.br:90/upcase/index_en.html in English and Brazilian Portuguese.

## 6. UPCASE Application and Evaluation

In order to evaluate the quality of the UPCASE method we conducted a series of evaluations as summarized in Table 7.

| Research question | Research design | Method for data collection | Method for data analysis |
|---|---|---|---|
| RQ1 Is the assessment method reliable? | Observed case studies | Questionnaire (UPCASE and on demographic data); Interview; Observation | Intraclass correlation analysis Weighted Kappa analysis |
| RQ2 Does the self-assessment method have good usability? | Observed case studies | Questionnaire (UPCASE and on demographic data); Interview; Observation | Analysis of the correct execution of the assessment process, the duration of the assessment and satisfaction of the participants. |
| RQ3 Is the method comprehensible? | Observed case studies | Questionnaire (UPCASE and on demographic data); Interview; Observation | Analysis of examples provided for each item of the questionnaire and correct execution of the assessment process. |
| RQ4 Is there evidence of internal | Remote unobserved case | | Cronbach' alpha analysis |

| consistency in the UPCASE questionnaire? | studies | | |
|---|---|---|---|

The characteristics evaluated by these research questions are:

- Reliability: the overall consistency of a measure, that means, if the same measuring process provides the same results. A measure is said to have a high reliability if it produces similar results under consistent conditions [89].

- Usability: the extent to which a product can be used by specific users to achieve specific goals with effectiveness, efficiency and satisfaction in a specific context of use [90]. In the context of this work, that means, whether the self-assessment process and the supplementary material (glossary and examples) may be used with efficiency, efficacy and satisfy the users.

- Comprehensibility: the extent to which a text as a whole is easy to understand [91]. In the context of this work, that means, the extent to which the items from the assessment questionnaire can be understood correctly.

- Internal consistency: the degree in which a set of items are measuring a single quality factor, i.e., the capability of a usability process [69].

## 6.1 Observed Case Studies

The evaluation of the reliability, usability and comprehensibility of the UPCASE method was done through case studies under the observation of a researcher [63]. The case studies consisted in the application of the UPCASE method in different small software organizations. The observed case studies were carried out in 4 small software organizations/projects in Brazil in October 2017, as shown in Table 8. The size of the assessment teams that participated in the case studies ranged from 12 to 47 persons. All small organizations develop web systems, including healthcare, governmental and marketing systems. Except for one, all participants of the assessment teams are either user interface designers or system analysts.

| Small organization | 1 | 2 | 3 | 4 |
|---|---|---|---|---|
| Number of employees | 30 | 12 | 47 | 18 |
| Domain / Platform | Governmental/Web | Health/Web | Health/Web | Other/Web |
| Number of participants | 4 (designers and system analysts) | 2 (designer and system analyst) | 7 (designers and system analysts) | 2 (front-end developers) |
| Assessment duration (minutes) | 20 | 60 | 80 | 43 |

**Table 8. Characteristics of the participating small organizations**

We invited the small organizations, via email, to participate in the study conducting a self-assessment of their usability process using UPCASE. The assessment meeting was supported using the UPCASE Tool. The participants performed the assessment autonomously, without interference of the researcher regarding the self-assessment process or the interpretation of the items of the questionnaire. During the assessment, a moderator presented the items of the assessment questionnaire. After obtaining a consensual response regarding each item, the moderator answered the respective question of the online questionnaire. During the assessment the observer recorded data on the usability and comprehensibility of the questionnaire and self-assessment process, including the duration of the assessment meeting, the usage of supplemental material, the correct interpretation of the questionnaire items, etc. After the assessment meeting the researcher asked the participants to present the work products cited as example during the meeting. Based on the information collected by the researcher, s/he also responded the UPCASE questionnaire separately for each of the participating organizations.

The duration of the assessment varied between 20 minutes and 80 minutes. In general, the duration of the assessments was considered appropriate (with an average of about one hour). The only assessment that lasted more than an hour was the one with a larger number of participants (seven participants). As expected, the number of participants in the assessment meeting influences the duration of the assessment. However, the factor that mostly influences the duration of the assessment is how critical the participants are and how much they discuss each questionnaire item and the number of examples they provide. Another reason that led to a longer duration of some assessment meetings was the attitude of the participants in trying to discuss how some of the items could be achieved, thus, initiating already an improvement of the usability process.

The result of the process assessment carried out in the case studies shows, that despite the variation of ratings obtained by the small organizations, all of them implement the usability process partially (attribute process = "P") (Table 9). In addition, it can be highlighted that in all four small organizations, the highest rated sub-process was UP3 (Produce design solutions) and the lowest rated was sub-process UP4 (Evaluate designs against requirements), indicating that the assessed small organizations have a greater capability in the development of design solutions, but do not yet adequately implement activities for evaluating the developed user interface designs.

| Small organization | 1 | | 2 | | 3 | | 4 | |
|---|---|---|---|---|---|---|---|---|
| | Assessment score | Attribute rating | Assessment score | Attribute rating | Assessment score | Attribute rating | Assessment score | Attribute rating |
| Usability process | 35 | P | 69 | P | 63 | P | 50 | P |
| UP1 | 35 | P | 67 | P | 50 | P | 50 | P |
| UP2 | 30 | P | 70 | P | 60 | P | 50 | P |
| UP3 | 60 | P | 90 | F | 100 | F | 60 | P |
| UP4 | 0 | N | 34 | P | 50 | P | 33,33 | P |

**Table 9. Assessment score and attribute rating of the assessed small organizations**

## RQ1 Is the assessment method reliable?

Table 10 presents the responses to the UPCASE questionnaire from the assessment team of the organization (T) and the observer (O) during the observed case studies. The answers "Full achieved", "Partially achieved" and "Not achieved" are respectively represented by "2", "1" and "0" .

| | UPCASE Questionnaire Item | Small organization | | | | | | | |
|---|---|---|---|---|---|---|---|---|---|
| | | 1 | | 2 | | 4 | | 4 | |
| | | T | O | T | O | T | O | T | O |
| 1 | Our team identifies and describes the purpose of the system. | 2 | 2 | 2 | 2 | 1 | 2 | 2 | 2 |
| 2 | Our team identifies stakeholders' expectations regarding the performance and behavior of the system. | 0 | 0 | 1 | 0 | 1 | 0 | 1 | 0 |
| 3 | Our team defines explicit statements of usability requirements based on the context analysis. | 0 | 0 | 1 | 0 | 1 | 0 | 0 | 0 |
| 4 | Our team identifies and describes the characteristics of the tasks the user performs in the system. | 1 | 1 | 1 | 1 | 1 | 1 | 0 | 0 |
| 5 | Our team identifies and describes the characteristics of the users. | 0 | 1 | 1 | 1 | 0 | 1 | 1 | 1 |
| 6 | Our team identifies and describes the organizational and social characteristics regarding the environment in which the system will be use. | 1 | 0 | 2 | 1 | 2 | 1 | 1 | 1 |
| 7 | Our team identifies and describes the characteristics of the device with which the users will interact to use the system. | 1 | 2 | 2 | 2 | 2 | 2 | 1 | 2 |
| 8 | Our team describes the physical environment characteristics in which the system will be use. | 0 | 2 | 1 | 2 | 1 | 2 | 2 | 2 |
| 9 | Our team analyzes the use cases in terms of its flow, navigation, main screens and constraints. | 1 | 1 | 2 | 1 | 2 | 1 | 1 | 0 |
| 10 | Our team analyzes a range of design options for each aspect of the system related to its use and its interface. | 1 | 1 | 1 | 2 | 2 | 1 | 1 | 1 |
| 11 | Our team applies existing usability knowledge (such as stakeholder requirements, usability guidelines) in the system design. | 1 | 1 | 2 | 2 | 2 | 2 | 2 | 1 |
| 12 | Our team specifies each aspect of the system related to its | 1 | 2 | 2 | 2 | 2 | 2 | 2 | 1 |
| | use and its interface. | | | | | | | | |
| 13 | Our team prototypes high-fidelity each component of the system interfaces. | 2 | 2 | 2 | 2 | 2 | 2 | 0 | 0 |
| 14 | Our team plans the prototypes and system evaluation. | 0 | 0 | 0 | 0 | 0 | 0 | 0 | 0 |
| 15 | Our team evaluates the usability of the prototypes. | 0 | 0 | 0 | 0 | 1 | 0 | 2 | 0 |
| 16 | Our team evaluates the system in order to check if it meets the usability requirements. | 0 | 0 | 2 | 1 | 0 | 0 | 0 | 0 |
| **Total** | | 34,375 | 46,875 | 68,75 | 59,375 | 62,5 | 53,125 | 50 | 34,375 |

**Table 10. Assessment responses**

The results of both assessments were compared with the objective of evaluating the reliability of the method. Reliability was analyzed through an intra-class correlation analysis [68] and a Weighted Kappa concordance analysis [67]. Intraclass correlation coefficient analysis is an estimate of the fraction of the total variability of measures due to variations between individuals. Intraclass correlation values in the range [0.4; 0.75] are considered satisfactory, and values greater than 0.75 are considered excellent [92]. Analyzing the intra-class correlation regarding the items of the UPCASE assessment questionnaire related to each process, the following coefficient were obtained, as presented in Table 11.

| Questionnaire section | Intraclass correlation coefficient |
|---|---|
| Items of UP 1 - Specify stakeholder and organizational requirements | -- |
| Items of UP 2 - Understand and specify the context of use | 0,5128 |
| Items of UP 3 - Produce design solutions | 0,7014 |
| Items of UP 4 - Evaluate designs against requirements | 0,4285 |
| **All sub-process items** | **0,5579** |

**Table 11. Intraclass correlation coefficient per questionnaire section**

The questionnaire items regarding the usability process 1 did not show variability in the results, which resulted in an undefined result. The intraclass correlation coefficient is calculated by the variance ratio. Since such values may be zero or negative, the use of this technique may lead to inconclusive results. On the other hand, the analysis of the questionnaire sections regarding the usability processes 2, 3, 4 and the usability process as a whole, presented a coefficient between 0.4 and 0.75, being considered a satisfactory correlation. This provides a first indication that the UPCASE assessment questionnaire presents a fair to good reliability when used in different moments to assess the same objects.

Analyzing the concordance between assessors regarding each item of the UPCASE assessment questionnaire, the Weighted Kappa coefficient was obtained, as presented in Table 12. Weighted Kappa measures the agreement between two raters, who each classify N items into C mutually exclusive categories. Weighted kappa allows to verify disagreements especially useful when codes are ordered [67]. Kappa over 0.75 is considered excellent, 0.40 to 0.75 fair to good, and below 0.40 as poor [93].

| Small organization | Weighted Kappa coefficient |
|---|---|
| 1 | 0,67033 |
| 2 | 0,351351 |
| 3 | 0,394958 |
| 4 | 0,461538 |

**Table 12. Weighted Kappa coefficient per SE**

The results indicate that the organizations 1 and 4 obtained coefficients higher than 0,4 and organizations 2 and 3 obtained a coefficient close to 0.4. Therefore, the Weighted Kappa coefficients indicate that the UPCASE questionnaire allows a reasonable agreement between different assessors.

### RQ2 Does the method has good usability?

An assessment method is considered to have good usability, if it is effective, efficient and satisfies the users, in this context the assessment team. During the case studies, all assessment teams were able to complete the self-assessment using the UPCASE method as planned without external assistance, which demonstrates that the method can be effective. The participating organizations were able to identify their usability process capability and improvement opportunities. However, we observed that in one case, participants did not follow the indication to turn over the assessment poker cards at once, showing them as soon as each participant had made his/her decision. As a result, we observed that the rating process was dominated by participants that showed their opinion first. We also observed, that with a larger number of participants in the assessment meeting it may take more time and more poker rounds to achieve consensus.

Considering that the assessment meetings lasted on average 50 minutes (with a maximum of 80 minutes), being an acceptable amount of time, the UPCASE method may be considered efficient.

When questioned in interviews after the assessment meeting, all participants indicated that they were satisfied with the way the assessment was conducted. One of the interviewees emphasized that the reason was the simplicity and quickness of the assessment process.

**RQ3 Is the method comprehensible?**

As the participants discussed each item of the questionnaire, the researcher observed whether they understood the items properly. This was perceived by the examples they provided to justify whether the respective practices were achieved or not. Based on these observations, we identified some items that need be better expressed and illustrated by more examples in order to prevent misinterpretations. The items that have turned out to be difficult to understand are presented in Table 13.

| Item | Difficulty | Small organization |
|------|-----------|--------------------|
| 2.Our team identifies and describes system performance and behavior requirements desired by the user. | Participants did not correctly understand the concept "performance". | 1 and 3 |
| 3.Our team defines explicit statements of usability requirements based on the context analysis. | Participants did not correctly understand the concept "usability requirement". | 2 and 3 |
| 4. Our team identifies and describes the characteristics of the tasks the user performs in the system. | Participants did not correctly understand the concept "task characteristics". | 2 |

**Table 13. Questionnaire items that were misunderstood**

In addition, the use of the term "describe" used by the items 1, 4, 5, 6, 7 and 8 was considered confusing by the participants of the organization 1 and 3, who thought that the verb "describe" requires documenting the activity. Consequently, this item may be considered "not achieved", even when the performance and behavior requirements are achieved but not documented. Therefore, the term "describe" has been substituted. Yet, considering the correct interpretation of the self-assessment process and the majority of the items of the assessment questionnaire, we can consider that the UPCASE is comprehensible.

## 6.2 Unobserved Case Studies

In addition to the observed case studies, we also conducted a series of remote unobserved case studies evaluating the internal consistency of the UPCASE method. To carry out these unobserved case studies, invitations were sent via email, social networks and forums to small organizations. The small organizations were asked to carry out a self-assessment of their usability process using the UPCASE method and tool (in English or Brazilian Portuguese). We collected the responses of the assessment questionnaire of each of the small organization that participated in the study, as well as demographic data (software platform and domain and number of employees). Data was collected from October, 2017 to January, 2018.

In total 36 organizations participated in the study. The participating organizations work in a variety of domains, especially in information technology, health and government sectors mostly developing web systems. The results of the usability process assessments showed that most of the assessed small organizations partially implement the usability process. In general, we observed, that the process with the highest capability is UP2: Understand and specify the context of use as shown in Figure 9.

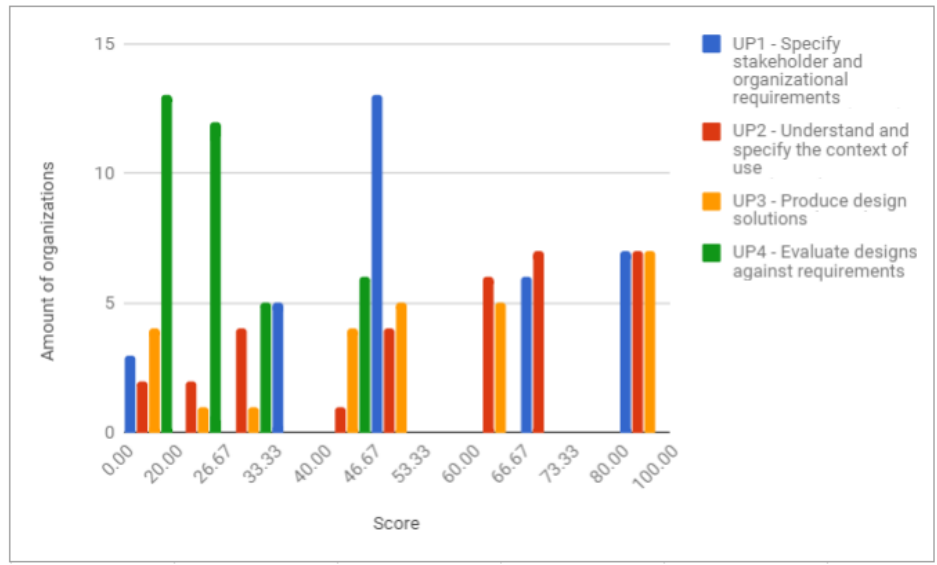

**Figure 9. Frequency of capability scores observed in the study**

### RQ4. Is there evidence of internal consistency of the UPCASE questionnaire?

The responses obtained from the case studies are used to evaluate the internal consistency of the UPCASE assessment questionnaire. To analyze the construct, we performed an analysis using Cronbach's alpha. Cronbach $\alpha$ values greater than 0.7 are considered acceptable [94], thus, indicating internal consistency of the measurement instrument.

Analyzing the items of the UPCASE assessment questionnaire, the value of Cronbach's alpha is satisfactory ($\alpha = .914$). This indicates strong evidence towards consistent answers, indicating the internal consistency of the measurement instrument. With the purpose of verifying if all items contribute to the reliability of the questionnaire, the Cronbach alpha coefficient was also calculated n times ("n" = number of items of the questionnaire), removing each time one of the questionnaire items (Table 14).

| Item removed | $\alpha$ | | Item removed | $\alpha$ | | Item removed | $\alpha$ | | Item removed | $\alpha$ |
|---|---|---|---|---|---|---|---|---|---|---|
| 1 | 0.910850877 | | 5 | 0.9077940077 | | 9 | 0.9067474106 | | 13 | 0.9075389559 |
| 2 | 0.9116298484 | | 6 | 0.906182051 | | 10 | 0.9068246905 | | 14 | 0.9071218217 |
| 3 | 0.9092369495 | | 7 | 0.907962094 | | 11 | 0.9040393173 | | 15 | 0.9093347639 |
| 4 | 0.9145593958 | | 8 | 0.9138603492 | | 12 | 0.9071273581 | | 16 | 0.9105609457 |

**Table 14. Alpha coefficient calculated with n-1 items**

As these values (Table 14) also are considered satisfactory, we can conclude that responses between the items are consistent and precise, indicating that the various items of the UPCASE assessment questionnaire purports to measure the same general construct.

## 6.3 Discussion

Based on existing SPCMMs and self-assessment methods, we systematically developed the UPCASE method in accordance with the specific characteristics of small organizations. Similar to most existing self-assessment methods we propose the use of a questionnaire for data collection. Yet, despite the wide adoption of questionnaires as a method for data collection, it does not come without shortcomings. The use of questionnaires may lead to unreliable responses (if the subject misinterprets a question) or/and lack completeness. Furthermore, questionnaires are typically answered individually, which makes it difficult to interact with the respondents in order to obtain further explanations on a given answer and/or to confirm the correct understanding of the items. In this context, interviews for collecting data may solve these issues. Interviews, however, also present some disadvantages, such as high cost (requiring people to conduct the interviews) and the collection of a small sample of data (as the size of the sample is limited to the number of interviewees) [46]. Furthermore, if conducted individually, inconsistencies and conflicts between the responses may have to be resolved later on. Therefore, we propose the combination of focus groups using a questionnaire for data collection as part of the UPCASE method. Focus groups are group interview that focus on a particular issue, product, service or topic and encompass the need for interactive discussion amongst participants [95]. Advantages of focus groups are that they allow the discussion of the indicators in order to achieve a consensus among a group of people. In comparison to individual interviews, focus groups can be more efficient capturing the opinion of a larger number of people at once as well as resolving conflicting responses immediately.

On the other hand, the realization of focus group meetings may lead to group effects with participants trying to dominate the discussion, while others may feel inhibited. Thus, some participants may publicly agree with others, while privately disagreeing. As a consequence, a reported consensus may be an opinion that not all participants really endorse or even disagree with [96]. To mitigate this risk, UPCASE uses an adaption of Planning poker [88], a consensus-based technique for estimating effort. We adapted this technique in order to assure the involvement of all participants in the decision-making process in order to increase the accuracy of the responses, while at the same time allowing the contribution of all participants. As result of the case studies, we also observed that in general the participating organizations were able to conduct the self-assessment process as expected, using only the material provided. Only one organization varied by not using the assessment poker technique. As a consequence, we observed that one participant dominated the conversation, while two other participants basically did not express their opinion. This shows how the use of assessment poker can help to gather the opinion of all participants.

The participants gave positive feedback with respect to the UPCASE process, especially the way the data is collected, enabling a rapid process assessment in a reasonable amount of time.

Having been conducted by employees of the small organizations, the results of the observed case studies also demonstrate that it may be possible to obtain valid assessment results using UPCASE. However, we also observed that the participants had difficulties in understanding some items of the assessment questionnaire (item 2: "Our team identifies and describes stakeholders' expectations regarding the performance and behavior of the system" , item 3 "Our team defines explicit statements of usability requirements based on the context analysis" and item 4 "Our team identifies and describes the characteristics of the tasks the user perform in the system"), which, thus, should be revised in order to facilitate comprehension. The analysis of the intra-class correlation coefficient with respect to the assessment questionnaire also showed an acceptable reproducibility of quantitative measurements made by different observers using the UPCASE method. Only questionnaire items with respect to UP1 "Specify stakeholder and user requirements" obtained inconclusive results and, thus, further cases studies are required to analyze this issue in more detail. The analysis of the weighted kappa coefficient indicated that the UPCASE method has a fair to good inter-rater agreement, which means that UPCASE is a fair to good method to assess the same object in different situations allowing agreement between assessors. Overall, the result indicated that the UPCASE questionnaire presents a fair degree of reliability when used in different moments to assess the same object. In terms of internal consistency, the results of the analysis indicate a satisfactory Cronbach alpha, which means that the items of the UPCASE questionnaire are measuring a single quality factor. These results, thus, provide a first indication that the UPCASE questionnaire is consistent and precise with respect to the assessment of the capability of the usability process of small organizations.

### 6.4 Threats to Validity

As in any research, there exist threats to its validity. Therefore, we identified potential threats and applied mitigation strategies in order to minimize their impact on the research results.

**External Validity.** Regarding external validity, a threat to the generalization of the results is related to the sample size and diversity of the data used for the evaluation [63]. With respect to sample size, we used data collected from observed and unobserved case studies, involving 4 and 32 small software organizations, respectively. In terms of statistical significance in general, this is a small sample size [97], yet, typical for empirical studies in software engineering research [98]. In addition, although inviting organizations on a national and international level, most organizations that participated in the studies are located in the state of Santa Catarina/Brazil. This shows that, although the study provides first explorative results, there exists a need for the realization of further case studies in different regions/countries as well as with larger sample sizes.

**Internal Validity.** Another issue refers to the correct choice of methods for conducting the data analysis. To minimize this threat, we performed a statistical evaluation based on the approach for the construction of measurement scales [69][69], and for assessing disagreement between respondents and measurement items correlation [67,68]. In respect to sample size and the user representativeness bias, we have used data collected from two series of case studies, involving a population of 4 and 32 organizations. In terms of statistical significance, these are small samples, yet, robust to estimate of the coefficient alpha (based on data from a total of 36 organizations) [99] and sufficient to indicate the generation of initial reliable results via intraclass correlation analysis and the concordance analysis (based on data from 4 organizations) [100].

**Construct Validity.** Threats to construction validity are related to the data collection instrument which may not contain the set of questions necessary to answer the assessment question: "what is the capability of the assessed usability process?". Therefore, we systematically developed the data collection instrument using the GQM approach, decomposing the assessment objective into indicators, operationalizing the data collection through a set of questionnaire items.

Considering that the interpretation of the collected data can be a main threat to the validity in a research, we carefully analyze the subjective data to evaluate the usability and comprehensibility of the UPCASE method. To reduce the risk of misinterpretation, the analysis was based on the triangulation of the data obtained from the case studies via questionnaires, interviews and observations.

## 7. Conclusion

Observing a lack of support for lightweight assessments of the usability process, we propose UPCASE, a method for self-assessing the capability of the usability process in small organizations. The method is based on the ISO/IEC 29110 and ISO/TR 18529 standard and is customized with respect to the specific requirements of small organizations. UPCASE incorporates a usability process reference model, a measurement framework, self-assessment process and an assessment model, including a questionnaire for data collection accompanied by supplemental material (such as a glossary and examples of work-products) and is supported through a web-based tool. The UPCASE method has been applied in 4 observed case studies and 32 remote unobserved case studies, in which small software organizations self-assessed their usability process. Results of these case studies provide a first indication that the method may be reliable. Feedback also indicates that the method is easy to use and understandable even for non-SPI experts. The results also show that with UPCASE small organizations may be able to assess their usability process in a quick and efficient way, identifying improvement opportunities in order to improve the usability of their software products.

**Acknowledgements**

We would like to thank all participants in the application for their valuable effort.

This work was supported by the CNPq (*Conselho Nacional de Desenvolvimento Científico e Tecnológico – www.cnpq.br*), an entity of the Brazilian government focused on scientific and technological development.